\documentstyle[12pt,psfig]{article}

\catcode`\@=11
\long\def\@makefntext#1{
\protect\noindent \hbox to 3.2pt {\hskip-.9pt
$^{{\ninerm\@thefnmark}}$\hfil}#1\hfill}                

\def\@makefnmark{\hbox to 0pt{$^{\@thefnmark}$\hss}}  

\def\ps@myheadings{\let\@mkboth\@gobbletwo
\def\@oddhead{\hbox{}
\rightmark\hfil\ninerm\thepage}
\def\@oddfoot{}\def\@evenhead{\ninerm\thepage\hfil
\leftmark\hbox{}}\def\@evenfoot{}
\def\sectionmark##1{}\def\subsectionmark##1{}}

\def\sectionc{\@startsection {section}{1}{\z@}{-3.5ex plus -1ex minus 
    -.2ex}{2.3ex plus .2ex}{\bf }}
\def\subsectionc{\@startsection{subsection}{2}{\z@}{-3.25ex plus -1ex minus 
   -.2ex}{1.5ex plus .2ex}{\it }}
\renewcommand{\section}[1]{\sectionc{#1}\hspace*{\parindent}}
\renewcommand{\subsection}[1]{\subsectionc{#1}\hspace*{\parindent}}
\newcounter{appendixc}
\newcounter{subappendixc}[appendixc]
\newcounter{subsubappendixc}[subappendixc]

\renewcommand{\appendix}[1] {\vspace*{0.6cm}
        \refstepcounter{appendixc}
        \setcounter{figure}{0}
        \setcounter{table}{0}
        \setcounter{equation}{0}
        \renewcommand{\thefigure}{\Alph{appendixc}.\arabic{figure}}
        \renewcommand{\thetable}{\Alph{appendixc}.\arabic{table}}
        \renewcommand{\theappendixc}{\Alph{appendixc}}
        \renewcommand{\theequation}{\Alph{appendixc}.\arabic{equation}}
        \noindent{\bf Appendix \theappendixc #1}\par\vspace*{0.4cm}}

\def\abstracts#1{{
        \centering{\begin{minipage}{13.2truecm}\footnotesize\baselineskip=13pt\noindent
        \parindent=0pt #1
        \end{minipage}}\par}}

\renewenvironment{thebibliography}[1]
        {\begin{list}{\arabic{enumi}.}
        {\usecounter{enumi}\setlength{\parsep}{0pt}
\setlength{\leftmargin 0.75cm}{\rightmargin 0pt}
         \setlength{\itemsep}{0pt} \settowidth
        {\labelwidth}{#1.}\sloppy}}{\end{list}}

\newcounter{itemlistc}
\newcounter{romanlistc}
\newcounter{alphlistc}
\newcounter{arabiclistc}

\newcommand{\fcaption}[1]{
        \refstepcounter{figure}
        \setbox\@tempboxa = \hbox{\footnotesize Figure~\thefigure. #1}
        \ifdim \wd\@tempboxa > 6in
           {\begin{center}
        \parbox{6in}{\footnotesize\baselineskip=13pt Figure~\thefigure. #1}
            \end{center}}
        \else
             {\begin{center}
             {\footnotesize Figure~\thefigure. #1}
              \end{center}}
        \fi}

\newcommand{\tcaption}[1]{
        \refstepcounter{table}
        \setbox\@tempboxa = \hbox{\footnotesize Table~\thetable. #1}
        \ifdim \wd\@tempboxa > 6in
           {\begin{center}
        \parbox{6in}{\footnotesize\baselineskip=13pt Table~\thetable. #1}
            \end{center}}
        \else
             {\begin{center}
             {\footnotesize Table~\thetable. #1}
              \end{center}}
        \fi}

\def\@citex[#1]#2{\if@filesw\immediate\write\@auxout
        {\string\citation{#2}}\fi
\def\@citea{}\@cite{\@for\@citeb:=#2\do
        {\@citea\def\@citea{,}\@ifundefined
        {b@\@citeb}{{\bf ?}\@warning
        {Citation `\@citeb' on page \thepage \space undefined}}
        {\csname b@\@citeb\endcsname}}}{#1}}

\newif\if@cghi
\def\cite{\@cghitrue\@ifnextchar [{\@tempswatrue
        \@citex}{\@tempswafalse\@citex[]}}
\def\citelow{\@cghifalse\@ifnextchar [{\@tempswatrue
        \@citex}{\@tempswafalse\@citex[]}}
\def\@cite#1#2{{$\null^{#1}$\if@tempswa\typeout
        {IJCGA warning: optional citation argument
        ignored: `#2'} \fi}}

 1
 1
 1

\font\ninerm=cmr9

\begin{document}

\centerline{\normalsize\bf Baryon Resonance Extraction from $\pi N$ Data using a 
Multichannel}
\centerline{\normalsize\bf Unitary Model}
\baselineskip=15pt

\vspace*{0.6cm}

\centerline{\footnotesize S.A. Dytman and T.P. Vrana}
\baselineskip=13pt
\centerline{\footnotesize\it Dept. of Physics and Astronomy, 
		University of Pittsburgh, Pittsburgh, PA 15260}
\vspace*{0.3cm}
\centerline{\footnotesize T.-S. H. Lee}
\baselineskip=13pt
\centerline{\footnotesize\it Physics Division, Argonne National Laboratory,
                             Argonne, IL 60439}

\vspace*{.3cm}
  
\vspace*{.6cm}
\abstracts{  
We present organization and solutions for a new analysis of all $\pi N$
elastic and the major inelastic channels to extract detailed characteristics
of the contributing baryon resonances.  This work is based on the work
of R. Cutkosky and collaborators at CMU about 20 years 
ago.  The model features analyticity at the amplitude level and
unitarity.  Results 
are similar to previous analyses for strongly excited states, but
can vary considerably from previous analyses when the states are
weak, the data is poor, or there is a strong model dependence.  We 
emphasize the S$_{11}(1535)$ resonance which has 
particularly strong model dependence.}

%

\normalsize\baselineskip=15pt
\section{Introduction} 
\label{se:Intro} 

	A primary goal in analyzing pion-nucleon elastic and inelastic data is
to ascertain the underlying resonant structure.  The properties of the N*
resonances are an important window into the behavior of strongly interacting
systems at large distance ($\sim$ 1 fm).  Many inelastic channels 
contribute roughly equally to the total
$\pi N$ total cross section.  A correct analysis should account for
all of them in establishing unitarity, matching all available data,
and including the proper threshold characteristics.

	The data must be decomposed into partial waves and separated into the 
various resonance contributions and their asymptotic 
channel excitation widths (e.g. the partial decay width into 
$\pi N$, $\eta N$, 
$\gamma N$, $\omega N$, $\rho N$, $\pi \Delta$, $\pi N^*(1440)$, and others) 
before the model calculations can be compared to data.  In the resonance 
region, the threshold
effects of asymptotic channels must be handled correctly to ensure a proper
identification of resonances.  This is particularly important for the 
S$_{11}(1535)$ where the $\eta N$ threshold comes just below the resonance
pole position.
	
	Resonance extraction requires a significant calculational effort 
and many 
articles have presented various ways to determine resonance parameters (masses, 
pole positions, and decay widths) from data.  The PDG mostly bases its 
recommendations on
older work by Cutkosky et al. (the Carnegie-Mellon Berkeley or 
CMB group)\cite{Cut79}
and H\"{o}hler et al. (the Karlsr\"{u}he-Helsinki or KH group)
\cite{KH80}, and 
more recent work by Manley and Saleski (or KSU) \cite{ManSal} and the VPI 
group.\cite{VPIpin} 
All these efforts use reaction data with $\pi N$ initial states.  All maintain 
unitarity, though the methods employed are quite different.  These models 
handle the multichannel character of the reactions in quite different ways.  
CMB and KSU use a formalism that allows for many channels while KH and VPI
focus 
on the $\pi N$ elastic channel by including a dummy channel to account for
all inelasticity.  For most strongly excited states, the four 
analyses tend to agree within expected errors on resonance masses and widths.  
A notable exception
is S$_{11}$(1535) where extracted full widths are 66 MeV\cite{VPIpin}, 
120$\pm$20 MeV\cite{KH80}, 151$\pm$27 MeV\cite{ManSal}, and 
270$\pm$50 MeV\cite{Cut79}.  This large variation is 
due to the close proximity of the resonance pole to the $\eta N$
threshold.

	This work presented here applies the CMB model\cite{Cut79} to 
$\pi N$ data with a large 
variety of final states.  This model emphasizes the proper treatment of all 
analytic features
that might be found in the complex energy plane.  The main publication 
where the CMB model was used was 
published in 1979.  Although they used both elastic and inelastic data, the 
elastic data was emphasized.  Significant analysis was required\cite{Kelly79} 
in order to represent the data in a model independent form.  

	Batinic et al.\cite{Bat} have applied the CMB model to more recent
data including three channels ($\pi N$, $\eta N$, and a dummy channel meant to 
represent 
$\pi \pi N$).  They use the KH80 model independent analysis of the $\pi N$
elastic data and the $\pi N \rightarrow \eta N$ data to produce a new fit
to the S$_{11}$ channel.  They determine a full width of the S$_{11}$(1535) 
to be 151 MeV.  

	We present a minimal account of the model and show 
the model dependence for the analysis of the data discussed above.  In this 
paper, we discuss some 
representative results.  We will provide a more complete list of baryon 
resonances and description of the model in a forthcoming paper\cite{TOMpin}.

\section{Features of the Model}
\label {se:model}
	The CMB model seeks a representation of the scattering T matrices
for many channels combining desirable properties of analyticity and
unitarity.  The phase space factors $\phi (s)$ (called the channel
propagators in the original paper) are calculated with a dispersion
relation which guarantees analyticity in the solutions.  This makes the
search for the actual pole of the T matrix in complex s space
possible.  Self energies are calculated for the coupling of each resonance
to asymptotic states as it propagates and are included via a Dyson
equation.  Since there are multiple open channels, we use the matrix form
of the Dyson equation to calculate the full resonance propagator G.
The self energies provide the required dressing
of the bare states to produce the physical states seen in experiments.  
Final states of two pions and a nucleon are included in
the phase space factors as quasi-two-body states with an appropriate
width.  

	A separable form for the T matrix is assumed.  Although this form
most easily allows reproduction of s-channel processes, additional 
nonresonant
processes are included in background.

\begin{figure}[t]
\begin{tabular}{cc}
\psfig{figure=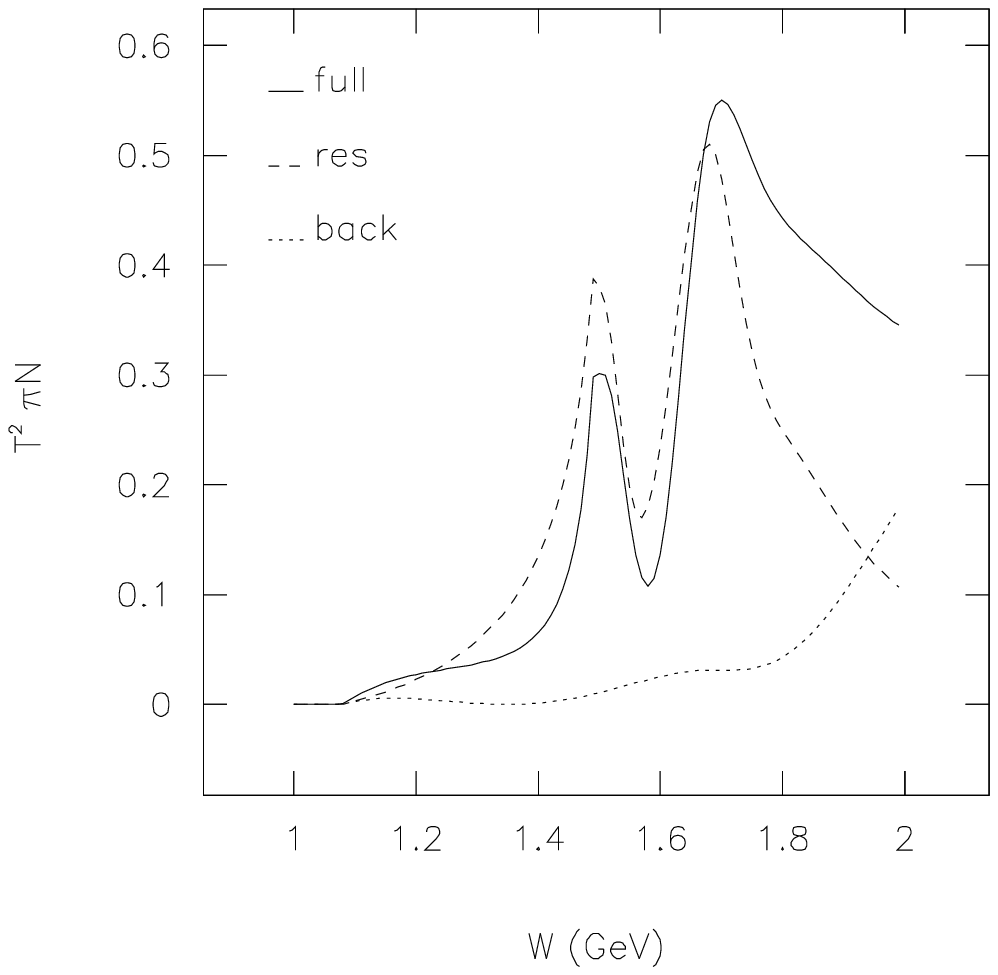,height=2.5in,width=2.8in}
 &
\psfig{figure=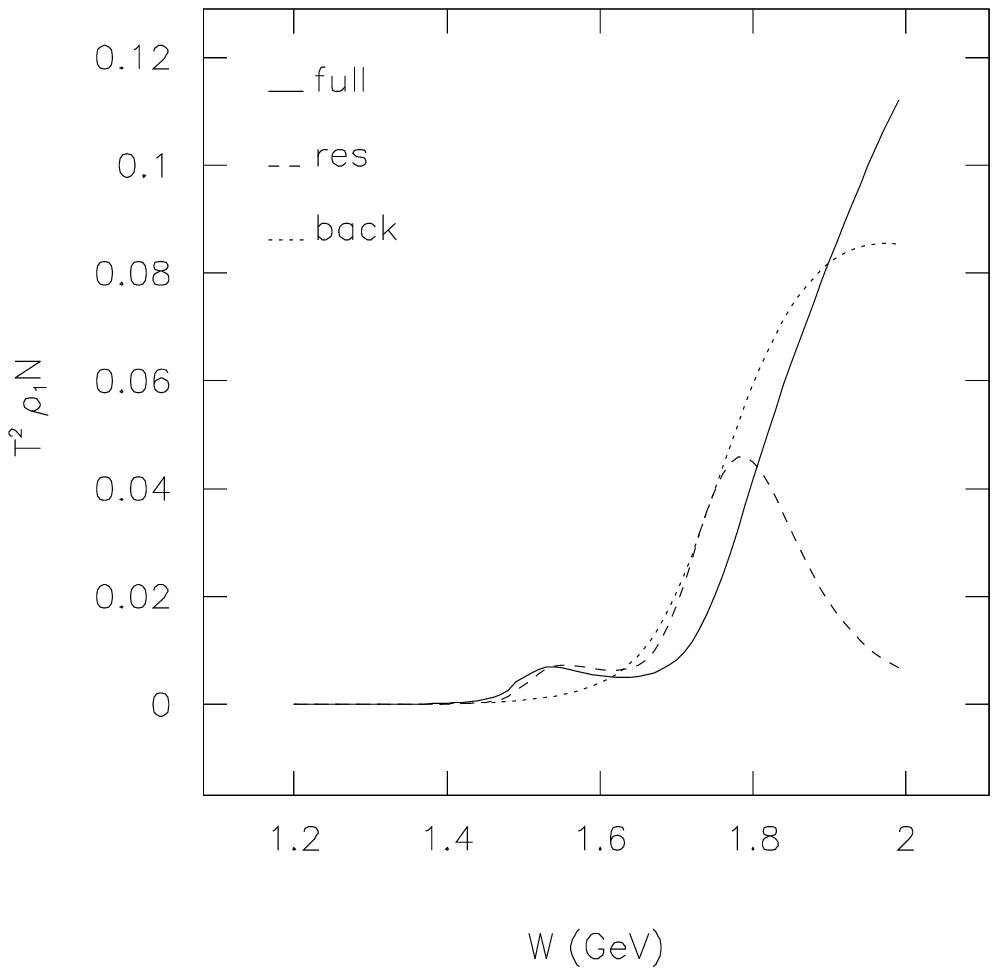,height=2.5in,width=2.8in}
\end{tabular}
\caption{Resonance-Background Separation for the $\pi$N and $\rho$N channels
	in the S$_{11}$ partial wave}
\label{fg:resnonres}
\end{figure}

	We have reproduced the CMB model~\cite{Cut79} and present fits to
modern data sets.  The same form factors and dispersion relations
are used.  Eight asymptotic channels are allowed to couple to the resonances
in each partial wave.  Two or three nonresonant terms are included in each 
partial wave.  These are represented as resonances at 
energies well below or well above
the resonance region.  Lacking a specific model for these processes, they have 
the same bare energy dependence (smooth) used for resonances, but the 
inelastic 
thresholds produce the appropriate analytic behavior for each nonresonant 
term.  We show the separation of resonant and nonresonant parts for the
magnitude of the T matrix element for elastic scattering and $\rho$
production in Figure~\ref{fg:resnonres}.

\section{Model Dependence in the S$_{11}$ Partial Wave}
\label{se:moddep}
	The variation between this and other models can best be seen in an
analysis of the data in the $\pi N$ S$_{11}$ channel.  Here, there are
2 strong resonances that have unusual features that are difficult to
describe.  The S$_{11}$(1535) has the $\eta N$ threshold close to the
peak of the resonance and the two states have significant overlap.
Data for both the $\pi N \rightarrow \pi N$ and $\pi N \rightarrow \eta N$
would be required for a high quality fit.  Presently, data for eta
production is of much less quality than for the elastic channel.  

	We have analyzed all available data for the S$_{11}$ channel with
the full CMB model and various approximations that simulate the models
employed by various other groups.  Results are shown in 
Tables~\ref{tb:mod1}~and~\ref{tb:mod2}.
Columns are labeled by the features of the model employed and the
results of that model.  All models are unitary, but this is done with
either a K matrix treatment or the full Dyson equation.  
To construct the K matrix model, we use the
truncated diagonal Dyson equation using just the first term.  Thus, the 
resonance coupling to other resonances
in the intermediate state is disabled.  The analytic properties are
turned off by setting the real part of $\phi(s)$ to zero; the 
imaginary part still contains the correct phase space.  We also compare
results obtained with a relativistic and nonrelativistic Breit-Wigner
shape for the bare resonance.  Finally, we used various data sets in
the fit since not all published models use data from all inelastic
channels.

\begin{table}[t]
\caption{Fitting results for the lowest energy resonance found in
the S$_{11}$ partial wave for various
models.  The model characteristics on the right four columns are
used to make the fits with results shown in the five columns on the
right.  See text for details.}
S$_{11}(1535)$\\
\begin{tabular*}{5.9in}{@{\extracolsep{\fill}}|ccccc|cccc|}
\hline\hline
 Mass & Width & $\pi$N & $\eta$N & $\pi\pi$N  & Unitarity & Disp. & Res & Channels \\
 MeV  & MeV   & \% & \% & \% &    &  Rel. & Type & in fit\\
\hline
1518 & 87 & 43 & 6 & 51 & K-Matrix & NO & NRBW & $\pi$N\\
1532 & 108 & 45 & 39 & 16 & K-Matrix & NO & NRBW & $\pi$N , $\eta$N\\
1535 & 126 & 42 & 44 & 14 & K-Matrix & NO & NRBW & ALL\\
1514 & 84 & 35 & 0 & 65 & K-Matrix & NO & RBW & $\pi$N\\
1533 & 110 & 44 & 40 & 16 & K-Matrix & NO & RBW & $\pi$N , $\eta$N\\
1534 & 125 & 42 & 43 & 15 & K-Matrix & NO & RBW & ALL\\
1531 & 72 & 16 & 62 & 22 & Dyson eq. & YES & RBW & $\pi$N\\
1526 & 114 & 36 & 41 & 23 & Dyson eq. & YES & RBW & $\pi$N , $\eta$N\\
1542 & 112 & 35 & 51 & 14 & Dyson eq. & YES & RBW & ALL\\
\hline
\end{tabular*}

\label{tb:mod1}
\end{table}
\begin{table}[t]
\caption{Same as Table~\ref{tb:mod1} for the S$_{11}(1650)$ Resonance.}
S$_{11}(1650)$\\
\begin{tabular*}{5.9in}{@{\extracolsep{\fill}}|ccccc|cccc|}
\hline\hline
 Mass & Width & $\pi$N & $\eta$N & $\pi\pi$N & Unitarity & Disp. & Res & Channels \\
 MeV  & MeV   & \% & \% & \% &    &  Rel. & Type & in fit\\
\hline
1645 & 233 & 35 & 49 & 16 & K-Matrix & NO & NRBW & $\pi$N\\
1689 & 225 & 67 & 31 & 2 & K-Matrix & NO & NRBW & $\pi$N , $\eta$N\\
1694 & 259 & 72 & 16 & 12 & K-Matrix & NO & NRBW & ALL\\
1682 & 161 & 78 & 5 & 17 & K-Matrix & NO & RBW & $\pi$N\\
1692 & 233 & 75 & 15 & 10 & K-Matrix & NO & RBW & $\pi$N , $\eta$N\\
1690 & 229 & 65 & 25 & 10 & K-Matrix & NO & RBW & ALL\\
1692 & 138 & 65 & 35 & 0 & Dyson eq. & YES & RBW & $\pi$N\\
1676 & 104 & 54 & 45 & 1 & Dyson eq. & YES & RBW & $\pi$N , $\eta$N\\
1689 & 202 & 74 & 6 & 20 & Dyson eq. & YES & RBW & ALL\\
\hline
\end{tabular*}
\label{tb:mod2}
\end{table}

	We first note that without the dispersion relation, the T matrices for 
the K-matrix model and the
model using the Dyson equation for the resonance propagator are
equivalent.  Therefore, only the K-matrix results without the dispersion
relation are given in the table.

	In general, the choice of relativistic vs. nonrelativistic shape for
the bare Breit-Wigner resonance does not have a strong influence for an
isolated resonance such as the 1535 MeV state.  However, the 1650 MeV 
state has a weaker signal (in part because of poorer data quality) and the
two shapes can produce larger differences.  Even there, the case where
all data is used (line 3 vs. line 6) has much less difference between RBW 
and NRBW.

	More important differences are found when comparing K-matrix vs.
Dyson results with the dispersion relation included (e.g. line 6 vs.
line 9).  The former is close to the model employed by Manley and Saleski.  
These two models have differences of about 10\% in the total width
and up to 50\% in the branching ratios.

	The most important deviation from the full result comes from the use
of a truncated data set.  For the 1535 MeV state, ignoring the interference
with the $\eta N$ final state causes the model to fit the Breit-Wigner
shape to the cusp at the $\eta N$ threshold.  The VPI work has a very
small width for the 1535 MeV state; although the $\eta N$ channel is 
mocked up, none of the actual data is used.  For even the full model, 
leaving out the $\pi NN$ final state data (such as was done by 
Batinic et al.\cite{Bat}) produces 20\% deviations in the branching ratios.  
We reproduce the updated results of the Batinic et al. paper\cite{Bat}.

\section{Results and Discussion}
\label{se:results}

	We have applied the CMB model to the $\pi N$ elastic T matrices of 
VPI\cite{VPIpin}, the inelastic T matrices of Manley et al~\cite{Manley84}, 
and our own partial wave analysis of the $\pi N \rightarrow \eta N$ data
(leaving out the controversial Brown et al. data\cite{Brown}).   
At this time, the VPI analysis\cite{Manley84} is the best available
information about the $\pi N \rightarrow \pi \pi N$ reactions.  A more 
objective analysis would use experimental data points
rather than the ``partial wave decomposed'' data we use in this paper.  A
reanalysis of the $\pi N \rightarrow\pi \pi N$ data is in progress at the
University of Pittsburgh; a more 
complete analysis can be presented when that is finished.

  The analysis presented here is a mixture of CMB\cite{Cut79} and
KSU\cite{ManSal} since we use the formalism of the former and a 
data set similar to that used by the latter.  Although the data set 
used here is very similar
to that used by KSU, they do not include the $\pi N \rightarrow \eta N$ data 
and use older elastic information.

	A partial list of resonances found in this analysis is given in 
Tables~\ref{tb:reslist} and \ref{tb:brlist} and compared to
the results of KSU, those of CMB and the latest recommended values 
given by PDG.  The number of states 
sought in each partial wave was the same as used by KSU\cite{ManSal}.  
We also show figures for a few representative cases.

\begin{figure}[t]
\begin{tabular}{cc}
\psfig{figure=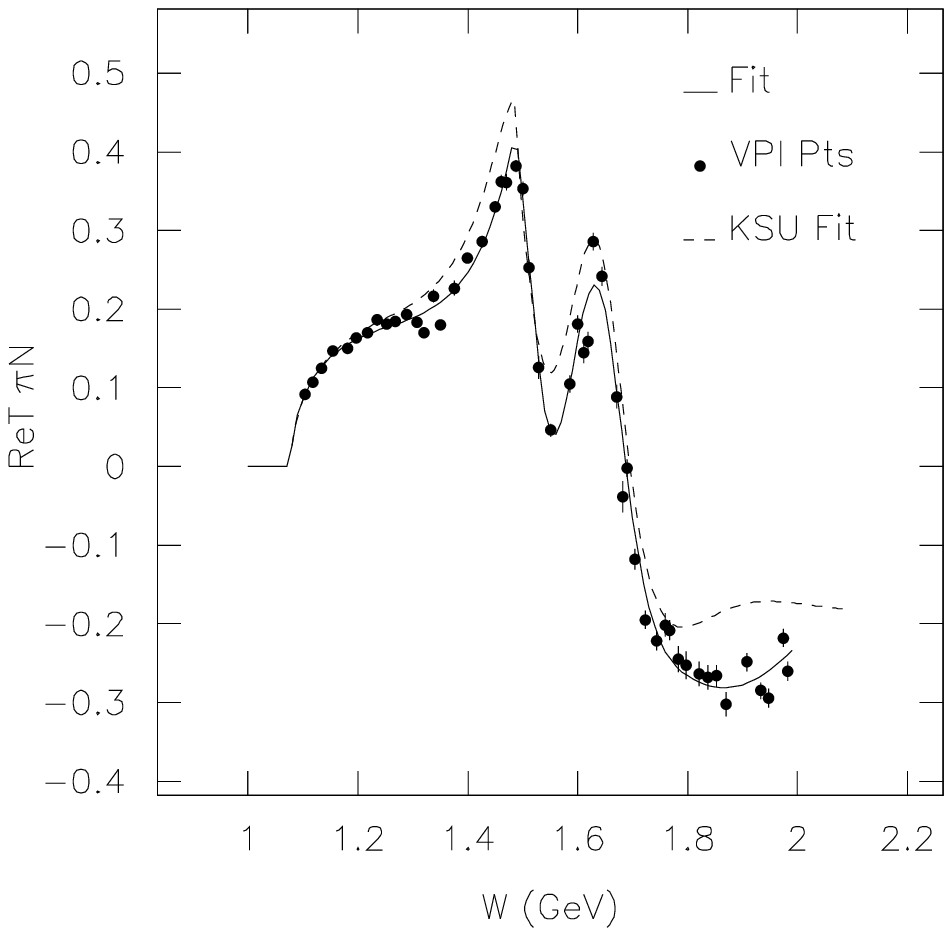,height=2.5in,width=2.8in}
 &
\psfig{figure=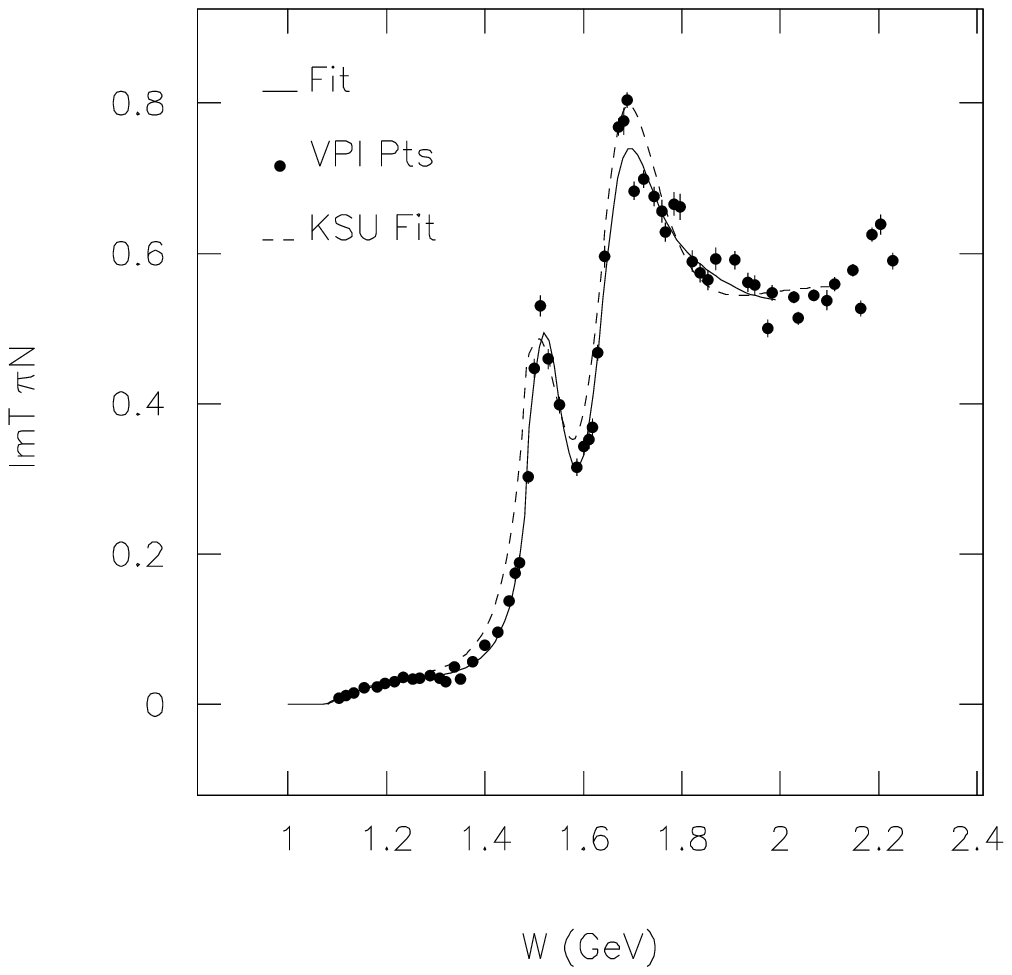,height=2.5in,width=2.8in}
\end{tabular}
\caption{S$_{11} \, \pi$N Elastic T-matrix Element}
\label{fg:s11pin}
\end{figure}

\begin{figure}
\begin{tabular}{cc}
\psfig{figure=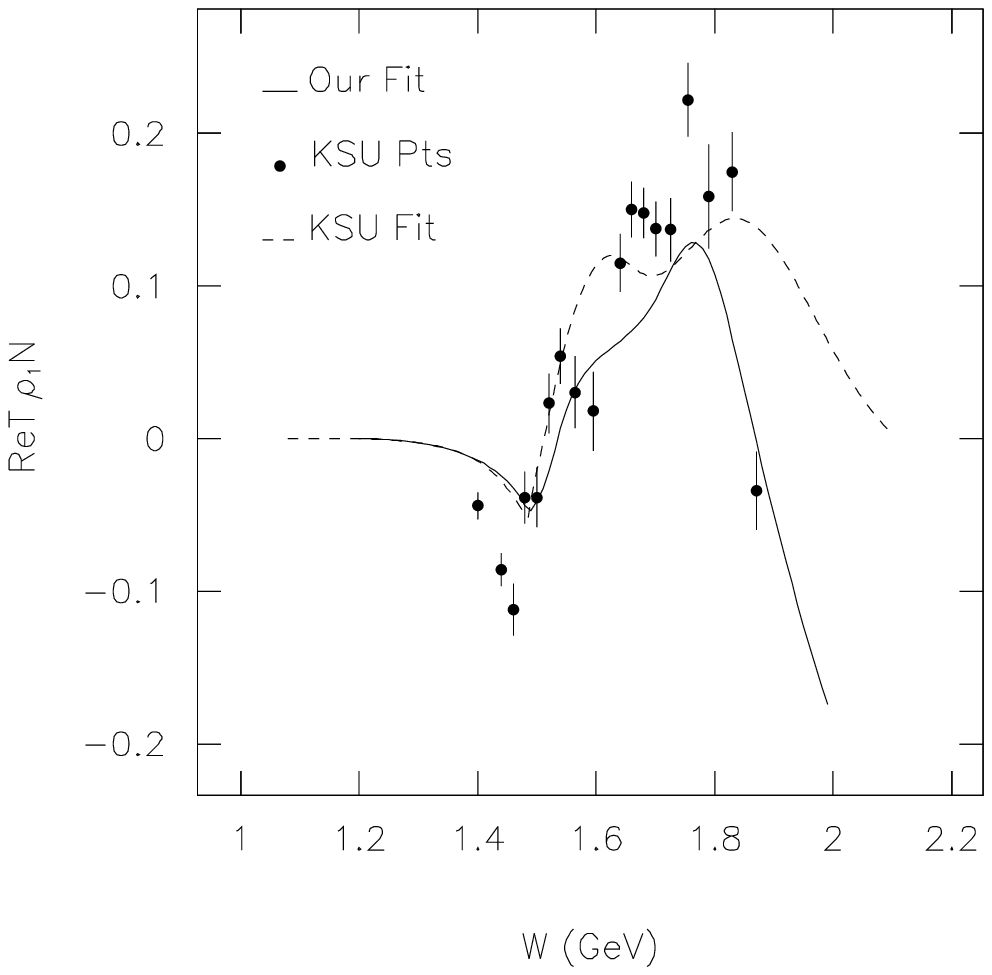,height=2.5in,width=2.8in}
 &
\psfig{figure=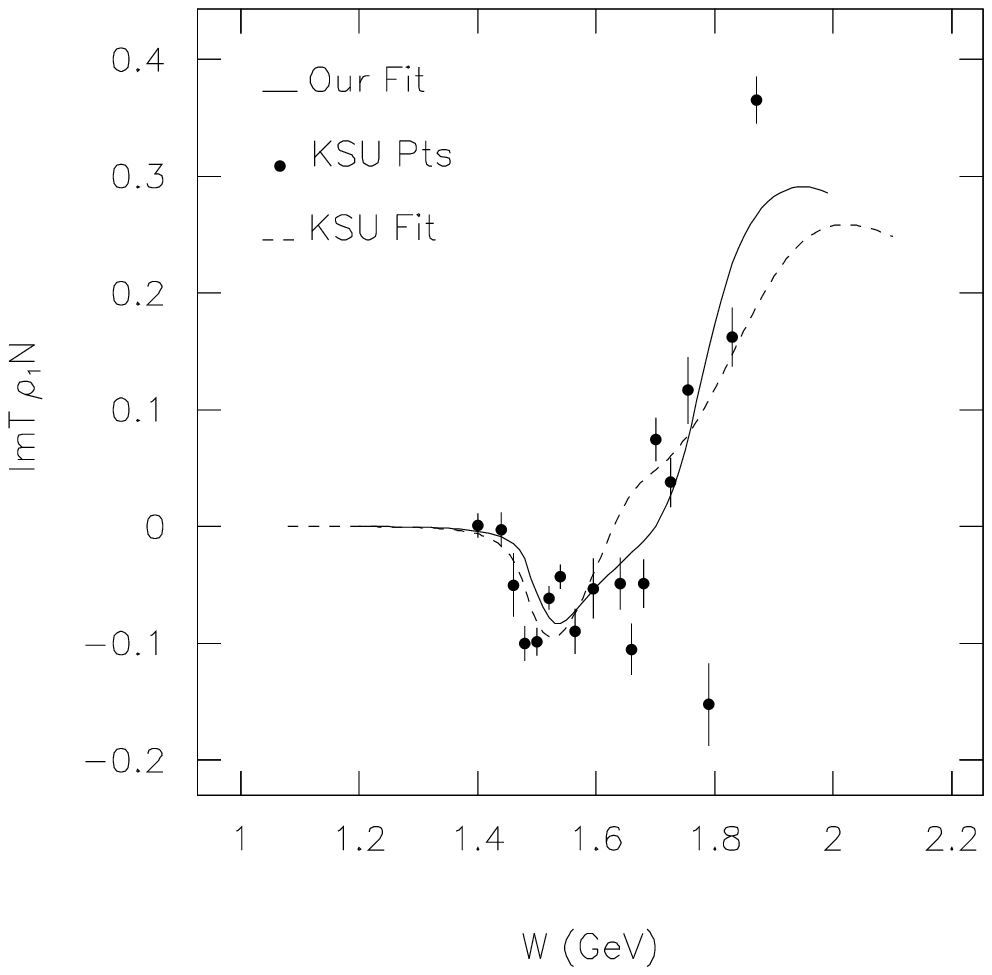,height=2.5in,width=2.8in}
\end{tabular}
\caption{S$_{11} \, \pi N \rightarrow \rho N$ T-matrix Element}
\label{fg:s11rhon}
\end{figure}

\begin{figure}[t]
\begin{tabular}{cc}
\psfig{figure=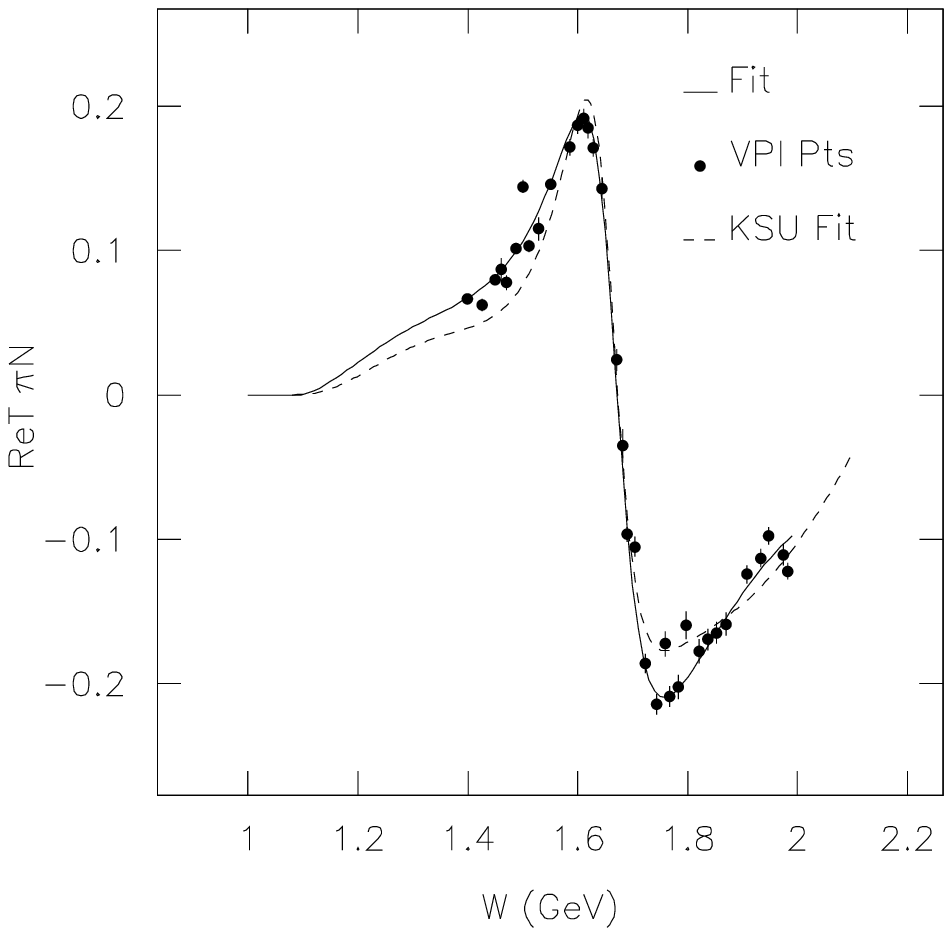,height=2.5in,width=2.8in}
 &
\psfig{figure=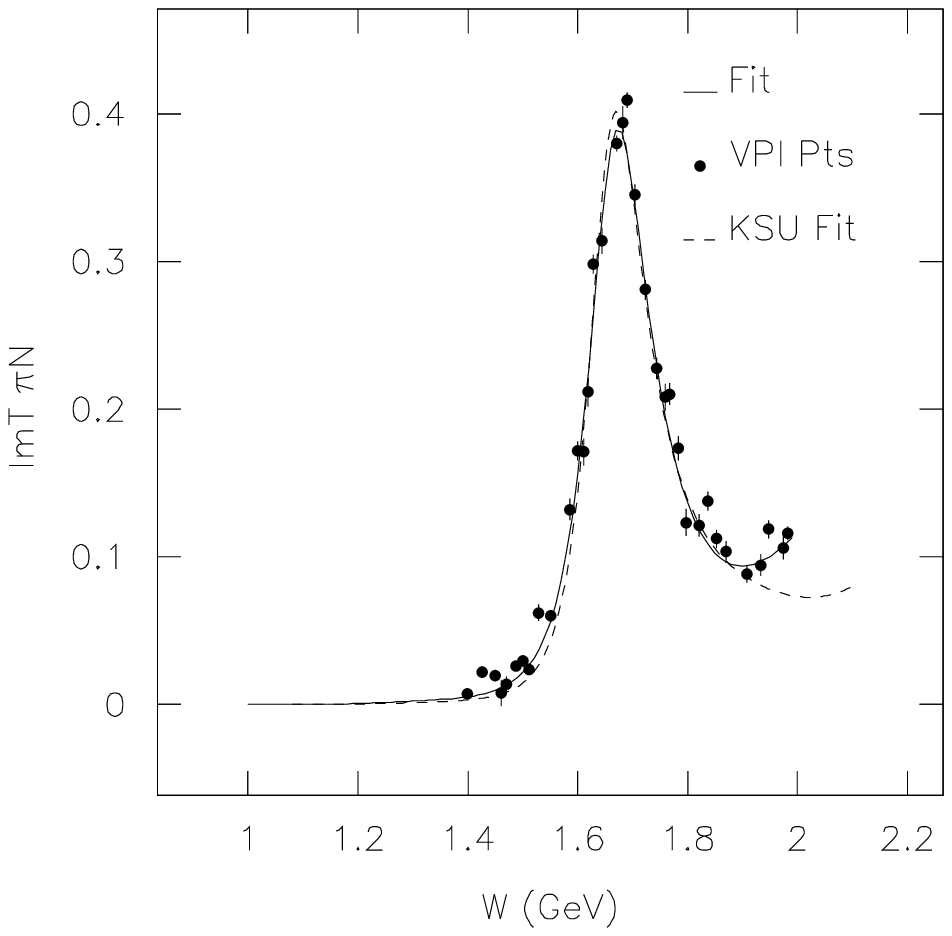,height=2.5in,width=2.8in}
\end{tabular}
\caption{D$_{15} \, \pi$N Elastic T-matrix Element}
\label{fg:d15pin}
\end{figure}

	The number of fit parameters depends on the number of 
resonances to be fit and
the inelastic couplings fit; for the S$_{11}$ partial wave (see 
Figs.~\ref{fg:s11pin} and \ref{fg:s11rhon}),
there are 3 resonances and we fit to the bare pole and the couplings to
7 channels for each.  There are also 2 subthreshold and 1 high energy `states'
used to simulate background.  Three parameters, a bare resonance energy 
and coupling strengths
to $\pi N$ and $\eta N$ are fit for each background pole.  Thus, there are 38 
parameters fit
in this partial wave.  For the D$_{15}$ partial wave 
(see Fig.~\ref{fg:d15pin}), only 1 resonance (with 
coupling to 4 channels) and 3 background poles were fit, a total of 14 real
parameters.  Since the elastic data is of much higher quality 
than the inelastic
data, the elastic data was weighted by a factor of 2 higher than the inelastic
in order to ensure a reasonable fit to the elastic data.  Although the elastic
data could be fit well, the inelastic data was not.  For elastic data,
typical values of $\chi^2/data point$ were 1.7 and 1.6 for the S$_{11}$ and
D$_{15}$ waves, respectively.  For the inelastic amplitudes, $\chi^2/data 
point$ values
were 9.2 and 22.2.  Although we get better fits to the elastic data than 
Manley and 
Saleski\cite{ManSal}, the inelastic fits are of similar quality in each 
analysis.  (No values of $\chi^2$ are given 
by KSU.)  However, the shapes of our inelastic T matrices are qualitatively 
different in many cases.

	For a complicated multiparameter fit, errors
are difficult to determine because correlations can be significant.  A Monte 
Carlo sampling technique was used.  
Resonance poles and physical quantities are found for many values 
of fit parameters, chosen according to the errors derived by MINUIT.   
Ten thousand such searches were made for each partial wave.  This method is 
similar to that employed
by KSU and similar errors result.  This error is almost entirely
statistical and only uses the diagonal errors from the partial wave
decompositions.  No systematic sources (e.g. due to the assumption of a
Breit-Wigner shape for the bare resonance) are included.  

	Strong isolated resonances that have a strong elastic coupling are  
fit well with all models.  The resonance parameters for these states,
e.g. the D$_{15}$(1675) and F$_{15}$(1680) masses and widths, tend to
have close agreement between previous results and the new results.

	The benefit of the multichannel analysis is readily apparent for states
with a very small elastic branching fraction.  For example, D$_{13}$(1700) is
not seen in the VPI~\cite{VPIpin} elastic analysis because the only sign of the
resonance in the elastic channel
is a shallow dip in the real part of the T matrix.  However,
there is a strong resonance signal in the $\pi N \rightarrow \pi \Delta$
T matrix.

	Agreement with the KSU analysis is mixed.  For the
cases where KSU differs significantly from the consensus of previous results, 
the S$_{31}$(1620) mass and the S$_{11}$(1650) elasticity, the new results
tend to agree with the older values.  Since the data sets used for the 
Pitt-Argonne and KSU analyses are very similar, any differences can most
likely be attributed to differences in the model employed.  To verify this,
we redid fits with the same elastic data as used by KSU and found
qualitatively similar results to those in the tables.  For F$_{15}$(1680), 
agreement is within errors for the largest decay branches.  
However, S$_{31}$(1620) is a strong state state where agreement
is not good.  For the S$_{31}(1920)$, the elastic``partial wave decomposed'' 
data has changed significantly from the set used by KSU to the set we
used.  The full width for KSU is 5 times larger than the new result.

\section{Conclusions}
\label{se:con}

	We present results for a new analysis of the {\it best available} 
$\pi N$
``partial wave decomposed'' data.  We apply a model~\cite{Cut79} that 
contains all the correct analytic
properties.  The results can differ significantly from previous analyses,
due to either the new data sets used in this work or due to model dependence.
A more complete discussion can be found in a forthcoming 
paper.~\cite{TOMpin}

	At present, the main limitation in any analysis is the data.  Improved 
$\pi N$ inelastic data is badly needed to generate high quality fits
required for objective results.

\hspace{.1in}

\noindent{\bf{Acknowledgments}}\newline
	We are grateful to Mark Manley and Dick Arndt for providing the 
``partial wave decomposed'' data used in this study.  We also wish to
thank Matt Mihalcin, Jay deMartino, and David Kokales for their
significant programming contributions to this work.

\begin{table}[ht]
\caption{Comparison of Resonance Masses and Widths for Selected Resonances}
\begin{tabular*}{5.9in}{@{\extracolsep{\fill}}ccccc}
\hline\hline
Resonance  &  Mass  & Width   &  Elasticity & Reference \\
           &  (MeV) & (MeV)   &  \%       &           \\
\hline
S$_{11}(1535)$ & 1542(15) & 112(30) & 35(5) & PITT-ARG\\
***** & 1534(7) & 151(27) & 51(5) & KSU\\
 & 1520-1555 & 100-250 & 35-55 & PDG\\
 & 1550(40) & 240(80) & 50(10) & CMB\\
\\
P$_{11}(1440)$ & 1479(9) & 490(18) & 72(3) & PITT-ARG\\
***** & 1462(10) & 391(34) & 69(3) & KSU\\
 & 1430-1470 & 250-450 & 60-70 & PDG\\
 & 1440(30) & 340(70) & 68(4) & CMB\\
\\
D$_{13}(1520)$ & 1518(13) & 124(25) & 63(4) & PITT-ARG\\
***** & 1524(4) & 124(8) & 59(3) & KSU\\
 & 1515-1530 & 110-135 & 50-60 & PDG\\
 & 1525(10) & 120(15) & 58(3) & CMB\\
\\
D$_{15}(1675)$ & 1685(13) & 131(26) & 35(4) & PITT-ARG\\
***** & 1676(2) & 159(7) & 47(2) & KSU\\
 & 1670-1685 & 140-180 & 40-50 & PDG\\
 & 1675(10) & 160(20) & 38(5) & CMB\\
\\
F$_{15}(1680)$ & 1679(11) & 128(23) & 69(4) & PITT-ARG\\
***** & 1684(4) & 139(8) & 70(3) & KSU\\
 & 1675-1690 & 120-140 & 60-70 & PDG\\
 & 1680(10) & 120(10) & 62(5) & CMB\\
\\
S$_{31}(1620)$ & 1617(10) & 143(19) & 45(3) & PITT-ARG\\
***** & 1672(7) & 154(37) & 9(2) & KSU\\
 & 1615-1675 & 120-180 & 20-30 & PDG\\
 & 1620(20) & 140(20) & 25(3) & CMB\\
\\
P$_{33}(1232)$ & 1234(2) & 112(3) & 100(1) & PITT-ARG\\
***** & 1231(1) & 118(4) & 100(0) & KSU\\
 & 1230-1234 & 115-125 & 98-100 & PDG\\
 & 1232(3) & 120(5) & 100(0) & CMB\\
\\
\hline
\end{tabular*}
\label{tb:reslist}
\end{table}

\begin{table}[ht]
\caption{Branching Ratios for Selected Resonances}
\begin{tabular*}{5.9in}[t]{@{\extracolsep{\fill}}ccccc}
\hline\hline
Resonance  &  Channel  & PITT-ARG   &  KSU  & PDG \\
\hline
S$_{11}(1535)$  & $\pi$N  &      35(5)  &      51(5)  & 35-55  \\
  & $\eta$N  &      51(6)  &      43(6)  & 30-55  \\
  & $\rho_1$N  &       2(6)  &       2(1)  & 0-4  \\
  & ($\rho_3$N)$_D$  &       0(0)  &       1(1)  &   \\
  & ($\pi\Delta$)$_D$  &       1(1)  &       0(0)  & 0-1  \\
  & ($\epsilon$N)$_P$  &       2(6)  &       1(1)  & 0-3  \\
  & $\pi$N$^*$(1440)  &      10(10)  &       2(2)  & 0-7  \\
\\
S$_{11}(1650)$  & $\pi$N  &      74(3)  &      89(7)  & 55-90  \\
  & $\eta$N  &       6(6)  &       3(5)  & 3-10  \\
  & $\rho_1$N  &       1(3)  &       0(0)  & 4-14  \\
  & ($\rho_3$N)$_D$  &      13(4)  &       3(2)  &   \\
  & ($\pi\Delta$)$_D$  &       2(5)  &       2(1)  & 3-7  \\
  & ($\epsilon$N)$_P$  &       1(4)  &       2(2)  & 0-4  \\
  & $\pi$N$^*$(1440)  &       3(5)  &       1(1)  & 0-5  \\
\\
D$_{15}(1675)$  & $\pi$N  &      35(4)  &      47(2)  & 40-50  \\
  & $\eta$N  &       0(0)  &   &   \\
  & $\rho_1$N  &       0(8)  &       0(0)  & 1-3  \\
  & ($\rho_3$N)$_D$  &       1(8)  &       0(0)  &   \\
  & ($\pi\Delta$)$_D$  &      63(4)  &      53(2)  & 50-60  \\
\\
F$_{15}(1680)$  & $\pi$N  &      69(4)  &      70(3)  & 60-70  \\
  & $\eta$N  &       0(0)  &   &   \\
  & ($\rho_3$N)$_F$  &       3(2)  &       2(1)  & 1-5  \\
  & ($\rho_3$N)$_P$  &       5(3)  &       5(3)  & 0-12  \\
  & ($\pi\Delta$)$_F$  &       1(3)  &       1(1)  & 0-2  \\
  & ($\pi\Delta$)$_P$  &      14(2)  &      10(3)  & 6-14  \\
  & ($\epsilon$N)$_D$  &       9(3)  &      12(3)  & 5-20  \\
\\
S$_{31}(1620)$  & $\pi$N  &      45(3)  &       9(2)  & 20-30  \\
  & $\rho_1$N  &      14(2)  &      25(6)  & 7-25  \\
  & ($\rho_3$N)$_D$  &       2(4)  &       4(3)  &   \\
  & ($\pi\Delta$)$_D$  &      39(2)  &      62(6)  & 30-60  \\
  & $\pi$N$^*$(1440)  &       0(6)  &   &   \\
\\
\hline
\end{tabular*}
\label{tb:brlist}
\end{table}

\end{document}